\documentclass[journal,twoside]{IEEEtran}

\usepackage[utf8]{inputenc} 
\usepackage[T1]{fontenc} 
\usepackage{amsmath}
\usepackage{amsfonts}
\usepackage{amsbsy}
\usepackage{amssymb}
\usepackage{bbm}
\usepackage[italicdiff]{physics}
\usepackage{proba}
\usepackage{enumerate}
\usepackage{cite}
\usepackage{wrapfig}
\usepackage{siunitx}
\usepackage{soul} 
\usepackage{times}
\usepackage{subcaption}
\usepackage{amsmath}
\usepackage{algorithm}
\usepackage[noend]{algpseudocode}
\usepackage{adjustbox}
\usepackage{xfrac}
\usepackage{url}
\usepackage{graphicx}
\usepackage{rotating}
\usepackage{mdframed}
\usepackage{cuted}
\usepackage{xcolor}
\usepackage{todonotes}
\usepackage{xr-hyper}
\usepackage[hidelinks]{hyperref} 
\usepackage{scalerel}
\usepackage{tikz}
\usepackage{textcomp}

\mdfdefinestyle{fig}{skipabove=3pt,leftmargin=0pt,innerleftmargin=3pt,innertopmargin=3pt,skipbelow=3pt,rightmargin=0pt,innerrightmargin=3pt,innerbottommargin=3pt}
\newcommand\cS{{\mathcal{S}}}

\title{\bf Machine Learning and Kalman Filtering for Nanomechanical Mass Spectrometry}

\author{Mete Erdogan, Nuri Berke Baytekin, Serhat Emre Coban, Alper~Demir
	\thanks{Authors are with the Department
		of Electrical Engineering, Ko\c{c} University, Istanbul 34450, Turkey.}
}	

\def\BibTeX{{\rm B\kern-.05em{\sc i\kern-.025em b}\kern-.08em
    T\kern-.1667em\lower.7ex\hbox{E}\kern-.125emX}}

\begin{document}

\maketitle
\flushbottom

\begin{abstract}
Nanomechanical resonant sensors are used in mass spectrometry via detection of resonance frequency jumps. There is a fundamental trade-off between detection speed and accuracy. Temporal and size resolution are limited by the resonator characteristics and noise. A Kalman filtering technique, augmented with maximum-likelihood estimation, was recently proposed as a Pareto optimal solution. We present enhancements and robust realizations for this technique, including a confidence boosted thresholding approach as well as machine learning for event detection. We describe learning techniques that are based on neural networks and boosted decision trees for temporal location and event size estimation. In the pure learning based approach that discards the Kalman filter, the raw data from the sensor are used in training a model for both location and size prediction. In the alternative approach that augments a Kalman filter, the event likelihood history is used in a binary classifier for event occurrence. Locations and sizes are predicted using maximum-likelihood, followed by a Kalman filter that continually improves the size estimate. We present detailed comparisons of the learning based schemes and the confidence boosted thresholding approach, and demonstrate robust performance for a practical realization. 
\end{abstract}

\section{Introduction}
\label{sec:intro}
\IEEEPARstart{N}{anomechanical} resonant sensors are used for the detection of nano-scale particles and atomic forces, with many applications in experimental physics, nano-engineering and molecular medicine~\cite{demir2020sensortradeoffs}. 
Nano-scale mass additions cause deviations in the resonance frequency, which can be detected and tracked using several schemes, such as the feedback-free (FF), the frequency-locked loop (FLL) and the self-sustaining oscillator (SSO) configurations~\cite{demir2020sensortradeoffs}.  
\subsection{Trade-off between speed and accuracy}
\label{sec:intro_tradeoff}
In the simplest FF tracking scheme, the response to a sudden change in the resonance frequency is limited by the mechanical time constant of the resonator, and can be modeled as the step response of a one-pole low pass filter with the transfer function $H_R(s) = \tfrac{1}{1+s\tau_r}$. Here, $\tau_r$ is the time constant of the resonator, and is inversely proportional to its quality factor. The step response, ignoring noise, is simply given by 
\begin{equation}
\label{eqn:expresp}
\Delta \omega_r(t) = \Delta \omega_e(1-e^{-t/\tau_r}), 
\end{equation}
for a sudden change $\Delta \omega_e$ in the resonance frequency. With modified FF, FLL and SSO schemes, one can speed up this response considerably but at the expense of degraded accuracy~\cite{bevsic2023resonance}. The degradation or improvement in the accuracy or speed stem from a change in the effective bandwidth of the system, which also determines its noise filtering characteristics. The noise sources in a nanomechanical sensor include the fundamental thermomechanical noise of the resonator, and noise generated in the transduction of the mechanical motion into an electrical signal and in the subsequent processing in the electrical domain~\cite{schmid2023}. There is a fundamental trade-off between speed and accuracy that can not be circumvented.

\subsection{Kalman filtering and likelihood based event detection}
\label{sec:intro_KF}
A Kalman filtering based technique  was recently proposed to detect and track resonance frequency changes~\cite{demirAKF2021}. This technique uses the raw output of the standard FF scheme operating in a slow speed but high accuracy point on the trade-off characteristics. Extremely fast predictions for both the temporal locations and also the sizes of events can be made using models for the system and streaming measured sensor data. Event size predictions, that are initially coarse, are then continually updated and improved. It was theoretically shown in~\cite{demirAKF2021} that this adaptive Kalman filtering and estimation based technique can achieve dynamic Pareto optimality with respect to speed versus accuracy and surpass the non-adaptive performance offered by the standard FF, FLL and SSO schemes.       

Kalman filtering combines streaming measurements from the actual device with the statistical predictions of a system model to obtain optimal estimates of the system state that can not be directly observed. However, the linear system model in the Kalman filter can not be used to predict jumps in the system state due to sudden resonance frequency changes. The Kalman filter is augmented with a maximum-likelihood estimation based event detection algorithm~\cite{Willsky1976}. In this technique, the differences between the model based predictions of the Kalman filter and the actual measurements from the nanomechanical device are continually processed with recursive equations to compute the {\it likelihoods} that a resonance frequency jump causing event may have occurred at time points in a sliding window with a prescribed duration. When the likelihoods exceed a certain predetermined threshold, it is decided that an event has occurred within the current time window. Then, the time point where the likelihood takes its maximum value is used as the prediction of the temporal location of the event. In addition, a maximum-likelihood based estimation for the size of the frequency jump is also made~\cite{Willsky1976}. After such event detection, the Kalman filter model and equations are updated and recursive processing is continued in tracking mode until next event occurrence. 

The Kalman filtering and estimation based technique outlined above was analyzed theoretically in~\cite{demirAKF2021}, and an equation for the monotonically decreasing variance of the frequency shift estimate as a function of elapsed time $t_{e}$ after event occurrence was derived, reproduced below
\begin{equation}
\label{eqn:varAKF}
\begin{split}
{\text{\bf var}}\pqty{\Delta\hat{y}} = \frac{\begin{array}{c}\textsf{Z}+ \sqrt{\cS_{y_{\textsf{\tiny th}}}t_{e}\:\textsf{Z}}\end{array}} {2\,t_{e}^{2}}
\\
\textsf{Z} = {\cS_{y_{\textsf{\tiny th}}}t_{e}+4\,\textsf{\small BW}_{\textsf{\tiny L}}\,\cS_{y_{\textsf{\tiny d}}} \tau_{\textsf{\tiny r}}^{2}}
\end{split}
\end{equation}
where $\Delta\hat{y}$ is the fractional (normalized by the nominal resonance frequency) frequency shift estimate, $\cS_{y_{\textsf{\tiny th}}}$ and $\cS_{y_{\textsf{\tiny d}}}$ are the spectral densities of the thermomechanical and transduction noise, and $\textsf{\small BW}_{\textsf{\tiny L}}$ is the effective bandwidth of the device. The equation above indicates that the frequency shift estimate eventually becomes perfect, whereas there is an upper limit to the accuracy that can be obtained with the non-adaptive, standard tracking schemes~\cite{demirAKF2021}.  

The recursive prediction and update equations for the tracking mode Kalman filter, as well as the ones for the maximum-likelihood based event detection algorithm~\cite{demirAKF2021,Willsky1976} can be used for an efficient, real-time DSP/FPGA based realization. While some guidelines are offered in~\cite{demirAKF2021} for such a practical implementation, no details were given. In particular,  
the choice of two crucial algorithmic parameters were not considered. The first one is the duration of the sliding time window in which an event is searched for. Windows with longer duration require more computations at every step of the algorithm but allow the detection of smaller events, since estimator variance decreases with elapsed time. Given $\Delta{y}_\textsf{\tiny min}$, size of the smallest event that needs to be detected, noise and system characteristics, \eqref{eqn:varAKF} can be used to determine the minimum window duration $M$ needed by setting it to the solution of  
\begin{equation}
\label{eqn:minsize}
{\text{\bf var}}\pqty{\Delta\hat{y}} = \left(\frac{\Delta{y}_\textsf{\tiny min}}{3}\right)^2
\end{equation}
for $t_{e}$. 
For reliable event detection, event size needs to be at least three times the estimate standard deviation. Otherwise, the event will be buried in noise and can not be discriminated.   

\subsection{This work and outline}
While the Kalman filtering based technique~\cite{demirAKF2021} was shown to surpass the performance of the standard schemes in theory, a fully fledged and practical implementation is needed to realize its potential. In this paper, we first present enhancements to this method. We consider the other crucial parameter, {\it i.e.,} the threshold value for the likelihoods that is used for the binary decision for event occurrence. Choosing the threshold value too large (small) results in too many missed events (false alarms). We present a technique for determining a proper threshold value that results in a robust algorithm for widely varying event sizes. Furthermore, we incorporate a confidence boosting scheme into the thresholding based decision to dramatically reduce the possibility of false alarms.  

We then present a fresh look into the event detection problem. We explore statistical machine learning (ML) techniques for the prediction of event temporal location and size. In the first technique that can be regarded as a {\it black-box ML} approach, we discard the Kalman filter and maximum-likelihood based estimation, and train an ML model using the raw data from the sensor. This ML model is used for the binary decision for event occurrence in a given window, as well as to predict both event location and size. In the second technique, we use an ML model as a replacement for the threshold test in Kalman filtering and maximum-likelihood estimation. In this {\it ML aided Kalman filtering} approach, the training is done with the likelihoods computed from the raw sensor data. In this case, the ML model is used only for the binary decision for event occurrence in a sliding window. When an event is detected as such, the location and the size in the window are predicted using the maximum-likelihood principle. We thoroughly investigate these two techniques, employing both neural networks and boosted decision trees for the ML model, and compare their performances with thresholding based event detection.

In Section \ref{sec:cbthrKF}, we describe the confidence boosted thresholding approach as well as a technique for determining the threshold value. In Section \ref{sec:bboxML}, we present the black-box ML based approach for event detection. The ML model for a binary classifier as a replacement for the threshold test in the Kalman filtering based technique is  discussed in Section \ref{sec:MLaidedKF}. Extensive comparisons of the proposed statistical learning based techniques and the thresholding based approach are presented in Section \ref{sec:results}. Conclusions are stated in Section \ref{sec:concl}.  

\section{Confidence boosted thresholding for \\ event detection}
\label{sec:cbthrKF}

In the maximum-likelihood based event detection algorithm~\cite{Willsky1976} that augments the Kalman filter, choosing the threshold value to detect small events could result in false alarms since the threshold may be exceeded momentarily due to a noise burst as opposed to an actual jump event~\cite{chow1976analytical}. While the threshold exceeding likelihood value due to a noise burst will be temporary, an actual jump event will result in persistently large values that grow as the likelihoods are updated in the subsequent steps~\cite{chow1976analytical}. Thus, event detection decision is taken only when the likelihoods exceed the threshold for $C$ consecutive steps, where $C$ is called the {\it confidence boosting parameter}, in order to reduce the possibility of false alarms, at the expense of increased latency for event detection~\cite{chow1976analytical}. The temporal location of the event is determined as the time point in the window where the likelihood takes its maximum value, whereas a separate maximum-likelihood based estimate of the event size is also computed at the same time~\cite{Willsky1976}. The accuracy of the size estimate improves with elapsed time after event occurrence, as given by \eqref{eqn:varAKF}. Thus, if event occurrence decision, and hence size estimation, is delayed as much as possible, this will further reduce the possibility of false alarms and increase the accuracy of both temporal event location and size predictions.  On the other hand, event occurrence decision  delay can not be larger than the window size. Thus, in addition to requiring the likelihoods to exceed the threshold for a number of consecutive time steps, we also require that event decision is taken only in the second half of the window.    

\begin{figure}
\centering
   \includegraphics[width=1.0\columnwidth]{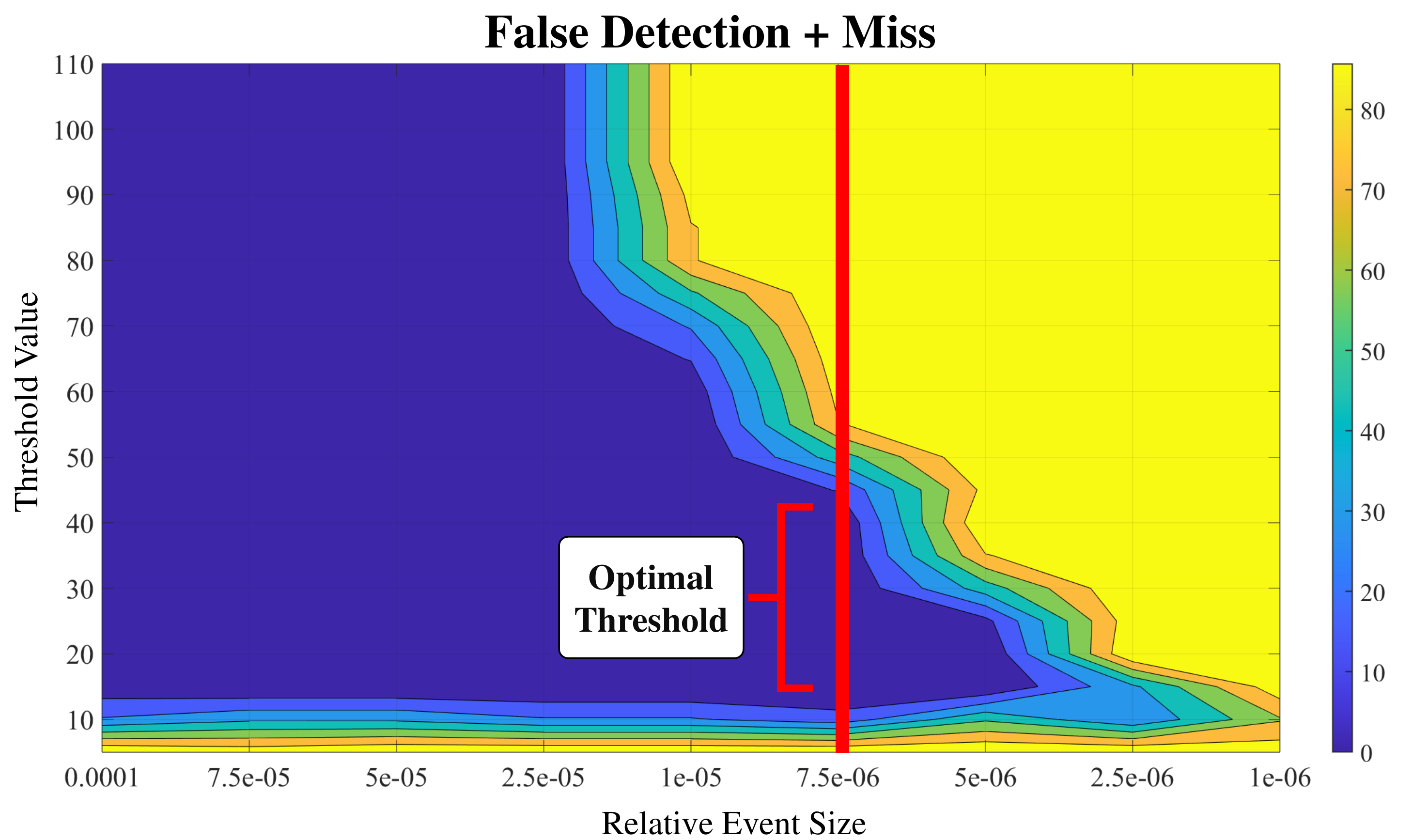}
\caption{Contour plot of False Detection + Miss with window size $M = 50$. Darkest (lightest) regions correspond to 0 (100)\% False Detection + Miss. The minimum detectable event size $7.5\times 10^{-6}$, and the optimal threshold range are indicated where both false detection and miss are 0\%.} 
\label{thcontour}
\end{figure}

\begin{figure}
    \centering
    \includegraphics[width=1.0\columnwidth]{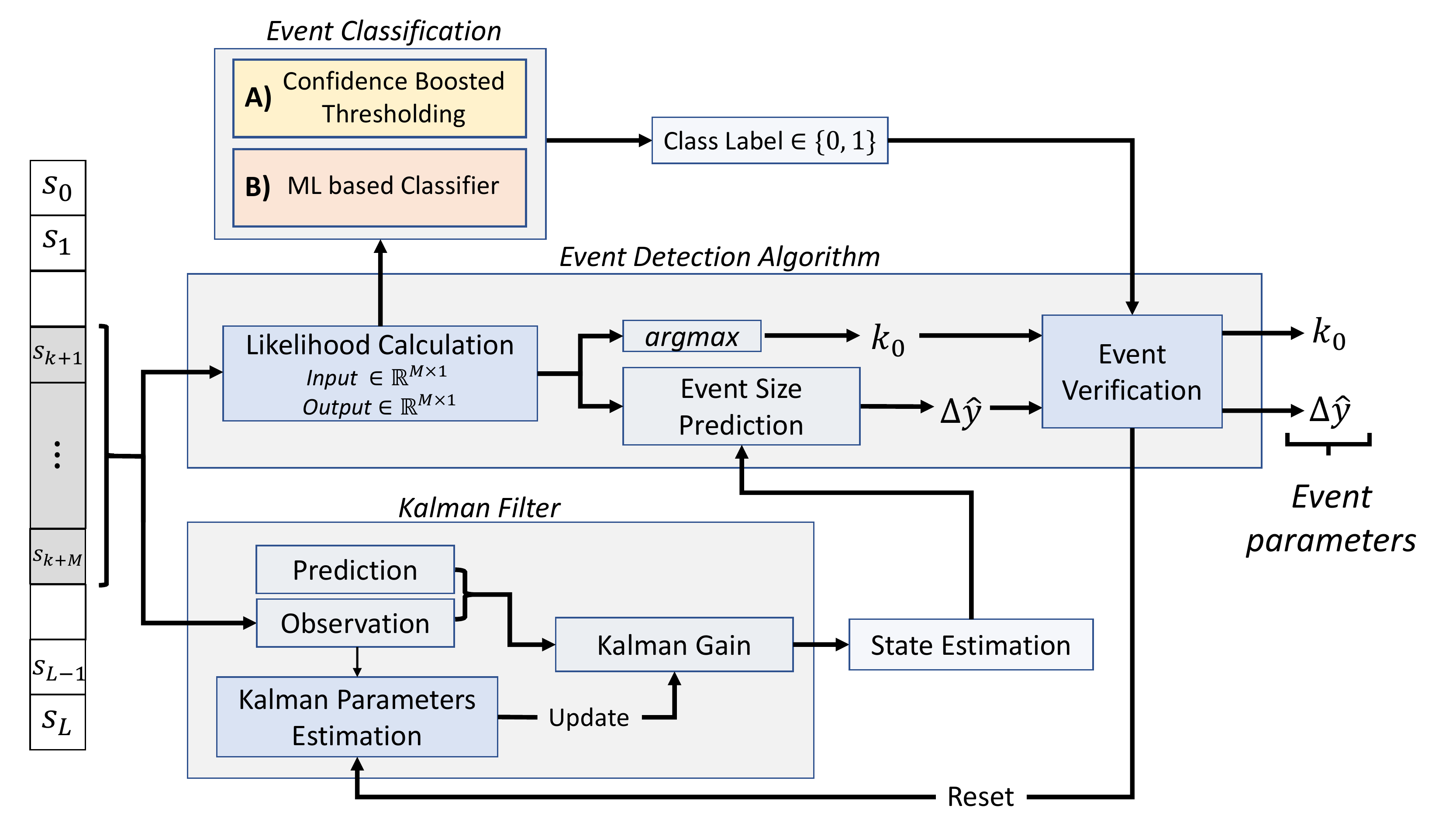}
   \caption{  Kalman filtering and maximum likelihood based event detection, with either Confidence Boosted Thresholding (A), or ML based Classifier (B) for event classification.}
    \label{fig:diag_kalman}
\end{figure}

The threshold value needs to be determined carefully, given noise and system characteristics and a range of event sizes. In fact, it suffices to determine the threshold value based on the size of the smallest event to be detected, which is also the determining factor for the minimum window size $M$ that is needed. Fig.~\ref{thcontour} shows that the threshold value optimized for the smallest event to be detected also works well for larger event sizes. We use a simulation based search algorithm, detailed in Algorithm~\ref{algorithmth}, to determine the optimal threshold value that corresponds to minimum false detection and missed event probabilities, and temporal location prediction error. A block diagram of the confidence boosted thresholding based Kalman filtering technique outlined above is shown in Fig.~\ref{fig:diag_kalman}.


\makeatletter
\def\BState{\State\hskip-\ALG@thistlm}
\makeatother
\begin{algorithm}
\caption{\\{\it th} (threshold) value is initialized to \textbf{LargeValue}, corresponding to 100\% misses. \textbf{r = Simulation(th, M)} performs simulations as described in Section~\ref{sec:results} using window size $M$ and threshold {\it th}. The output {\it r} has fields {\it fd} (false detection), {\it m} (miss) and {\it loc\_err} (location error), averaged over the simulations with minimum detectable event size for a given $M$. $\delta$ sets the stop condition of the threshold search.}
{\small
\begin{algorithmic}[1]
\Procedure{SearchOptimalThreshold}{$M, \delta$}
    \State $th \leftarrow \textbf{LargeValue}$
    \State $th\_low \leftarrow 0$
    \State $r \leftarrow \textbf{Simulation(th, M)}$
    \State $th\_prev \leftarrow th$
    \State $r\_prev \leftarrow r$

    \While{True}
        \If{$th-th\_low < \delta$}
        \State $\textbf{break}$
        \EndIf
        \State $r \leftarrow \textbf{Simulation(th, M)}$
        \If{$\Call{IsBetterThreshold}{r, r\_prev}$}
            \State $th\_prev \leftarrow th$
            \State $th \leftarrow th-\lfloor(th-th\_low)/2\rfloor$
            \State $r\_prev \leftarrow r$
        \Else
            \State $th\_low \leftarrow th$
            \State $th \leftarrow th\_prev-\lfloor(th\_prev-th\_low)/2\rfloor$
        \EndIf
    \EndWhile  \label{roy's loop}
    \State \textbf{return} th\_prev
\EndProcedure
\\
\Procedure{IsBetterThreshold}{$r_1,r_2$}
    \If{$r_1.m$ = 100\%}
        \State \textbf{return} True
    \ElsIf{$r_1.fd = r_2.fd$ and $r_1.m = r_2.m$}
        \If{$r_1.loc\_err \leq r_2.loc\_err$}
        \State \textbf{return} True
        \Else
        \State \textbf{return} False
        \EndIf
    \ElsIf{$r_1.fd \leq r_2.fd$ and $r_1.m \leq r_2.m)$}
        \State \textbf{return} True
    \Else
        \State \textbf{return} False
    \EndIf
\EndProcedure
\end{algorithmic}
}
\label{algorithmth}
\end{algorithm}

\section{Black-box ML for event detection}
\label{sec:bboxML}
Kalman filtering can be regarded as a form of ``physics-informed'' ML\cite{karniadakis2021physics} for making predictions about dynamical systems affected by noise, since it uses a model for the system. On the other hand, modern ML techniques are very good at solving many classification and estimation problems. Hence, we also explore the effectiveness of a black-box ML approach for the resonance frequency jump detection and estimation problem we address in this paper. In this end-to-end ML scheme, we discard the Kalman filter and maximum-likelihood based estimation and use the raw sensor data both for training and inference.      

The raw sensor output for a frequency jump that occurs at $t=t_0$ is essentially an exponential response time-shifted by $t_0$, but affected by noise. In the black-box approach, the aim is to detect jump events in real-time immediately after they occur, long before the response settles to its steady-state, and estimate both the time of occurrence $t_0$ as well as the size $\Delta \omega_e$ of the jump, all based on ML models. Our approach is two-pronged: (i)~Binary classifier for event occurrence in a sliding time window, followed by (ii)~regression based estimation of event location and also event size once it is decided that an event has occurred. The binary classifier needs to perform well in the presence of noise and for a wide range of event sizes, minimizing the probability of false alarms as well as missed events. 

\begin{figure}
    \centering
    \includegraphics[width=1.0\columnwidth]{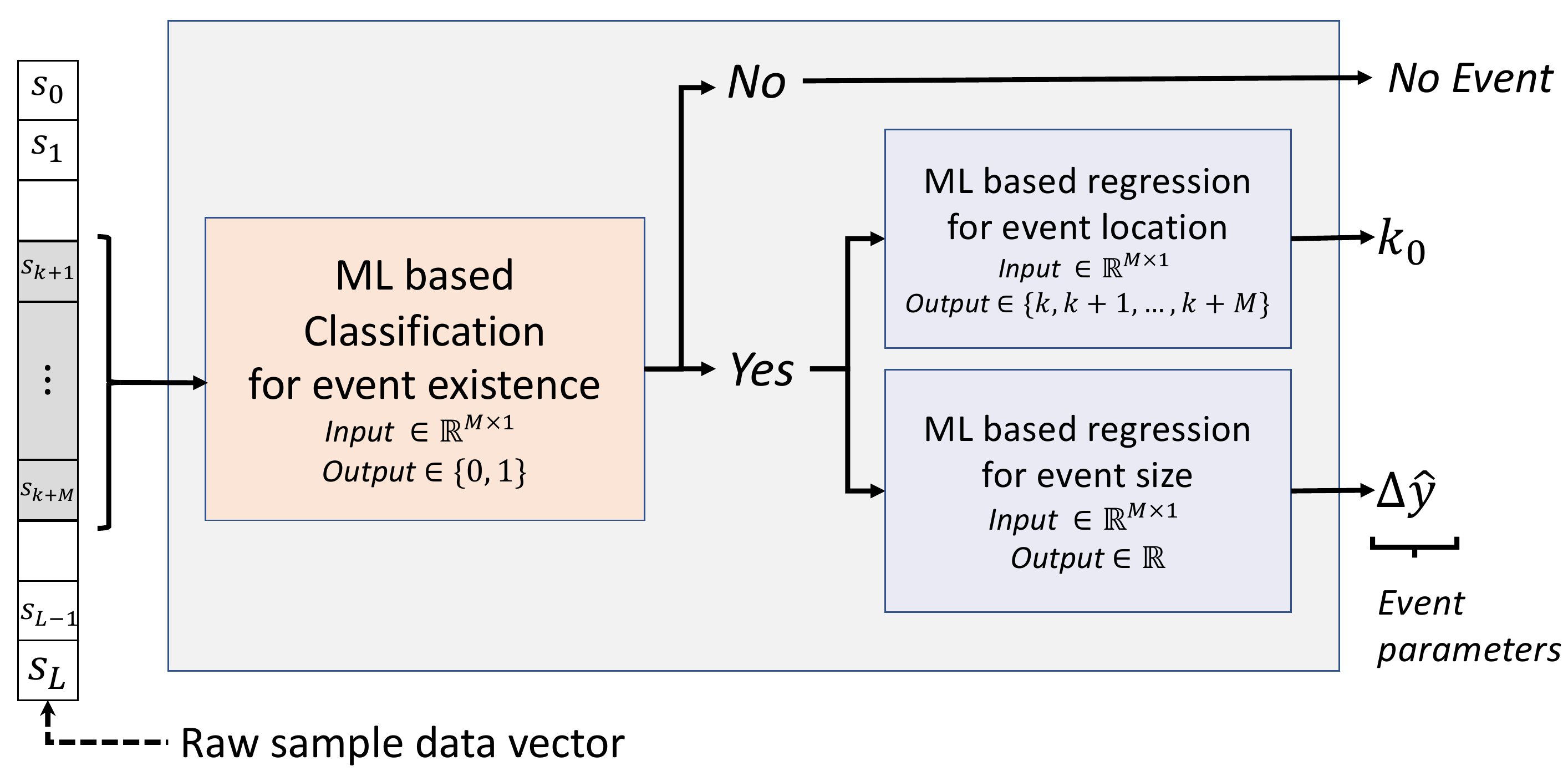}
    \caption{  Black-box ML for event detection}
    \label{fig:diag_brute}
\end{figure}

We assume availability of training data in the form of sampled time-series for the raw sensor output exhibiting fluctuations due to noise and containing jump events. We divide this time-series into stride-1 windows with size $M$, where there is at most one event in a given window. Each window is labeled with a binary value for event occurrence, and in the case of event presence, with event location within the window and event size.  Due to stride-1 windowing, the training data contains a balanced number of windows with(out) events, even if the original events in the time-series are not frequent. We use the labeled data to first train a binary classification network for event occurrence in a given window. In addition, we train two other networks to perform regression for event location and size. Alternatively, multi-output regression can be used where one network predicts both location and size. Such multi-output networks are known to perform better due to generalization by solving two related problems at once~\cite{multi-output-regression,multi-output-regression2}. We consider three architectures for the classification and regression networks: (i)~Regular fully-connected neural network (RFNN), (ii)~convolutional neural network (CNN), (iii)~XGBoost scheme~\cite{XGBoost}.  A CNN could result in more effective training due to the highly correlated nature of the training data~\cite{CNN}. XGBoost is a boosted tree scheme that works well, even without extensive hyper-parameter tuning, by choosing a suitable maximum tree depth parameter. Results and comparisons for the black-box approach outlined above and shown in Fig.~\ref{fig:diag_brute} will be presented in Section \ref{sec:results}. 
\section{ML-aided maximum-likelihood event detection}
\label{sec:MLaidedKF}
We explore whether the performance of maximum-likelihood based event detection can be improved by replacing the simple threshold test with an ML based binary classifier. In this approach, we keep the Kalman filter and the associated likelihood computations. We use the likelihoods in a sliding time window as the feature set and training data for the binary classifier, as opposed to the unprocessed raw sensor data that is used in the black-box approach discussed in Section \ref{sec:bboxML}. When the outcome of the binary classifier is positive, the temporal location of the event within the window is determined by the time point corresponding to the maximum likelihood value. Event size prediction based on maximum-likelihood rule is also retained. The binary classifier is invoked after each time sample of raw data from the sensor is processed to update the history of the likelihoods in a sliding window.      

The likelihood time-series is processed and organized to generate the training data for the binary classifier. For each time-series of likelihoods with length $N$, there is exactly one event occurring in one of the data samples. This sequence of likelihoods is used to generate $N-M$ sequences of length $M$ using stride-1 windowing. The ratio of windows that contain an event to the total is given by $\tfrac{M}{N}$. We choose $N\approx 20\,M$ to have a balanced set of windows with(out) an event. We filter the sequences where the event is located very close to the boundaries of the window. We generate a number of training windows as such, that is inversely proportional to window size $M$, so that the training data size for varying window sizes are equivalent. A window of length $M$ is labeled as 1 (0) for event occurrence if it (does not) contains a sample with an event. Binary classifier is to determine whether an event has occurred in any of the samples of a window with consecutive $M$ likelihoods. In order for event detection to be versatile, the data set contains events in a range of sizes as determined by the requirements of the application.   

We consider three approaches for binary classification: (i)~RFNN, (ii)~XGBoost scheme~\cite{XGBoost}, (iii)~RusBoost scheme~\cite{RUSboost}. To improve performance, event presence is declared only if the binary classifier outcome is positive for $C$ consecutive time steps, as in the confidence boosted thresholding approach. The size estimate, computed after event decision, will be further improved by the tracking mode Kalman filter, whereas the location estimate is frozen once event occurrence is declared and the Kalman filter is reset. Results and comparisons for the ML aided Kalman filtering approach outlined above and shown in Fig.~\ref{fig:diag_kalman} will be presented in Section \ref{sec:results}.
\flushbottom

\section{Results}
\label{sec:results}
\subsection{Generation of synthetic raw sensor data}
Due to the unavailability of real sensor data with labeled events, the mathematical sensor model given in \cite{demirAKF2021} was used for synthetic data generation. The predictions of this model were shown to match experimental measurements extremely well~\cite{sadeghi2020frequency}. To emulate the thermomechanical and transduction noise sources with given specifications, a Gaussian random number generator is used. Frequency jump events are inserted during synthetic data generation, with appropriate labeling for event occurrence, temporal location and size. Events sizes in a desired range with a log-uniform distribution are generated. 

The generation and organization of raw sensor training data for the black-box ML and the likelihood based ML approaches are slightly different. For black-box ML, two separate data sets for the classification and the regression networks are generated. The training data set for the classification network consists of sessions with 10\% containing exactly one event, while the rest do not contain any events. However, in the data set for the regression network, all sessions contain an event. The sessions are organized into time-series' with length $M$, same as the window size used during inference. The event location is uniformly distributed in the window except for the first and last  $\tfrac{\text{M}}{10}$ samples. In the case of the likelihood based ML approach, only a classification network is trained. The session length $N = 20\,M$  is chosen to be larger than the window size $M$. The event location is uniformly distributed in the window except for the first and last $M$ samples. The raw sensor data that is generated and organized as described above are used directly in the black-box ML approach, whereas in likelihood based ML, the raw sensor data are first processed with the Kalman filter. Then for each session of length  $20M$, the computed likelihoods are organized into time-series with length $M$ and stride-1 windowing as described in Section \ref{sec:MLaidedKF}, resulting in a total of $19M$ time-series, where  $M$ of them contain an event and the rest do not. 

\subsection{Specifications for black-box ML}

For binary classification of event occurrence, 4 fully connected layers with 100, 50, 20 and 1 (output) neurons per layer in an RFNN are used. A ReLU activation function is chosen between the layers except for the last layer, where a sigmoid function is preferred instead. For combined event location and size prediction, we use a single regression network with a mean squared error loss function. Regression network has 5 fully connected layers with 500, 250, 100, 10 and 2 (output) neurons per layer. Adam Optimizer~\cite{adamoptimizer} is used in the training of both the classification and the regression networks. 

For CNN, we use a structure where 2 consequent convolutional layers are followed by a max-pooling layer. We use this structure 3 times in a row, resulting in a total of 6 convolutional and 3 max-pooling layers, where each convolution has 16 kernels and a padding that results in no spatial dimension reduction. We append 2 fully connected layers with 50, 10 and 1 (output) neurons for the binary classification network, and 50, 10 and 2 neurons for the multi-output regression network. 

For the XGBoost approach, an XGBoost based classifier is trained first. Then, using the {\it multi-output-regressor method} of the scikit-learn library~\cite{scikit}, an XGBoost regressor with two outputs for event size and location is created. While multi-output regression with one XGBoost model is convenient for implementation, it is equivalent to two separate XGBoost models each with one output, unlike the case with multi-output neural network models that result in improved performance.  The maximum depth of the tree was chosen as 11 in both the classification and the regression models.

\subsection{Specifications for likelihood based classifier}

An RFNN with 4 fully connected layers with 50, 50, 10 and 1 (output) neurons per layer
are used.  Again, ReLU activation function is used for each layer except for the last one, where a sigmoid function is used. Fully connected layers with non-linear activation functions are deemed sufficient for the binary classification task based on likelihoods,  as opposed to using more complex architectures. The model loss function used is {\it binary cross entropy}, as is common when training a binary classifier with the Adam Optimizer in the stochastic gradient descent algorithm. As the window size $M$ is a variable in the architecture, the learning decay rate in the optimizer is adjusted to sweep a large range of values from $10^{-3}$ to $10^{-8}$ for better convergence. Such networks were created and tested using both the Tensorflow and Keras libraries in Python, as well as with the Deep Learning Toolbox of $\text{MATLAB}\textsuperscript{\textregistered}$~\cite{matlab}. 
Another approach based on XGBoost that uses decision trees to perform binary classification is also implemented, with a maximum tree depth of 11. The tree based RUSBoost~\cite{RUSboost} classifier was also considered, which peforms well when the training data is imbalanced.

\subsection{Event detection algorithms in action} 

\begin{figure}
    \centering
    \includegraphics[width=1.0\columnwidth]{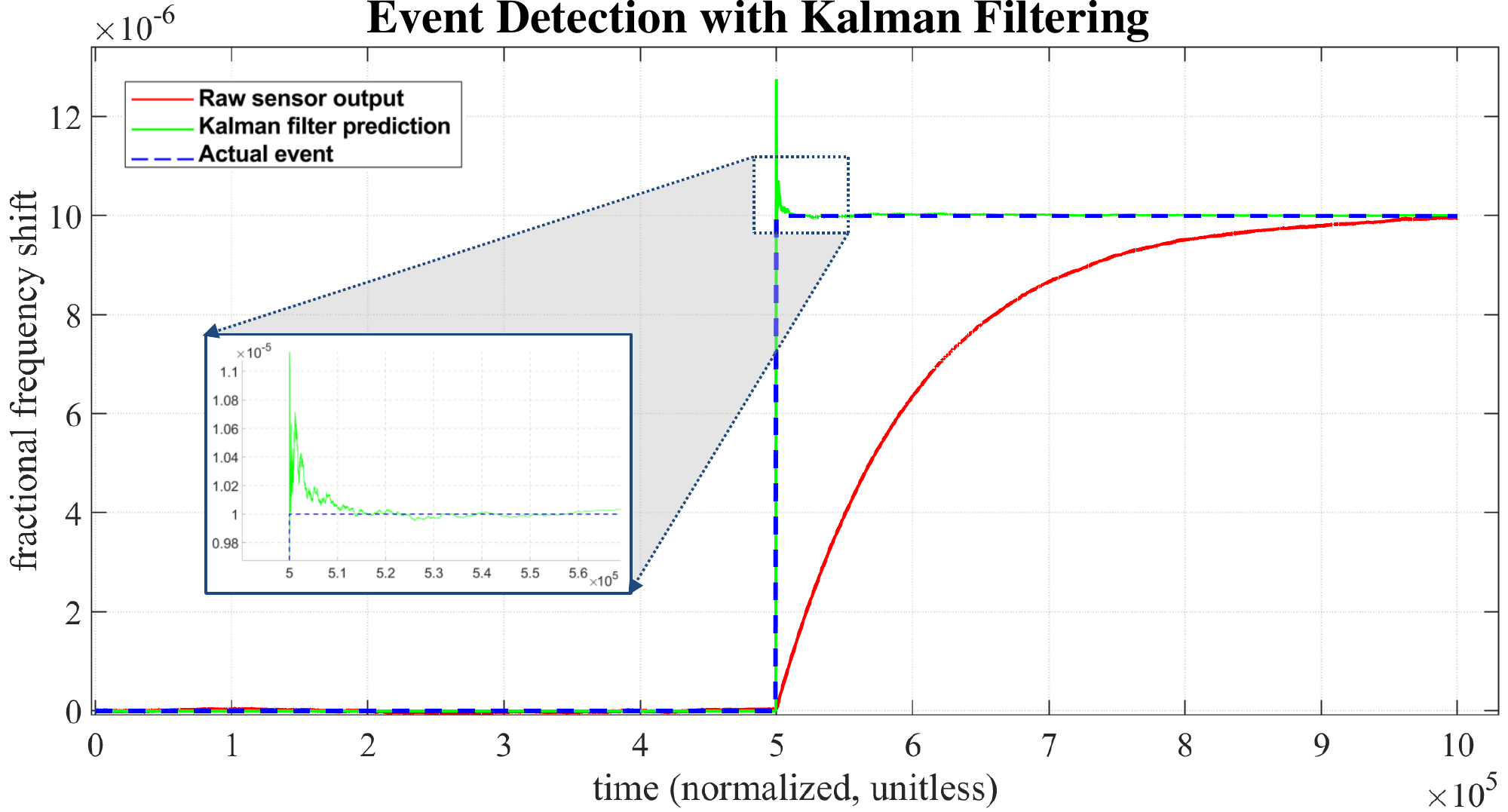}
    \caption{  Event prediction with ML-aided Kalman filtering.}
    \label{fig:one_event}
\end{figure}

The event detection algorithm (with ML aided Kalman filtering) is illustrated in Fig.~\ref{fig:one_event}, showing a single ground-truth frequency jump event, and the predictions by the proposed  algorithms. The raw sensor output that settles to the steady-state with a very large time constant is also shown. It can be observed that events can be detected very soon after they occur with high accuracy. Thresholding based binary classifier for event detection performs in the same manner as the ML aided method. Event size prediction is performed with the Kalman filter and the maximum-likelihood based rule in both cases.

\subsection{Performance evaluation and comparison}

\subsubsection*{Performance criteria and methodology}
We assess the performances of the thresholding technique and the various ML schemes described above by computing the accuracy of the classifications and regressions when they are used on synthetic and labeled raw sensor output in the form of a time-series (not included in training). 
We compare the various variants (RFNN, CNN, XGBoost) of both black-box ML (Section~\ref{sec:bboxML}) and likelihood based ML that augments a Kalman filter (Section~\ref{sec:MLaidedKF}), as well as confidence boosted thresholding described in Section~\ref{sec:cbthrKF}. These schemes are compared with each other with respect to event classification accuracy quantified in terms of false alarms and missed events, event temporal location and size estimation accuracy, event detection latency, computational cost of processing and inference per new sample of raw sensor output. Two window lengths and a range of event sizes are considered. The performance of  the confidence boosting scheme is demonstrated. Various errors and accuracy metrics are computed by running the schemes on 1000 different sets of synthetic data. Random number generators are used for noise emulation, event occurrence, location and size in the generation of test data.       
We quantify event temporal location estimation accuracy by computing the difference between mean estimated and exact location, and also root-mean squared error (RMSE) of location estimate. Event size prediction is performed via a regression model in black-box ML, but using a maximum-likelihood criterion in likelihood based ML and the thresholding approach that augments a Kalman filter. While size prediction is a one-shot regression look-up that is then frozen in black-box ML, it is continually updated and improved by the Kalman filter in the likelihood based approaches after event occurrence decision. In the latter case, the size estimate accuracy is a function of detection latency. Thus, we report size estimate accuracy both at event occurrence decision and also after the estimate is improved by the Kalman filter for ${10\,M}$ time steps. 

In confidence boosting, {\it i.e.,} requiring a number of consecutive positive decisions for declaring event occurrence, we choose the repetition count, {\it i.e.,}, the confidence, to be a fraction of the window size. Therefore, mean detection latency comparisons for various schemes will be presented for a fixed window size and a fixed confidence.   

In order to assess the computational cost of all of the techniques, we estimate the run-time of each one for every new sample from the streaming raw sensor output. We report run-time comparisons by varying the window size $M$ and event size $\Delta \omega_e$. The run-times are indicative of the performance one would obtain in a DSP/FPGA based realization. 

\begin{figure}
    \centering
    \includegraphics[width=1.0\columnwidth]{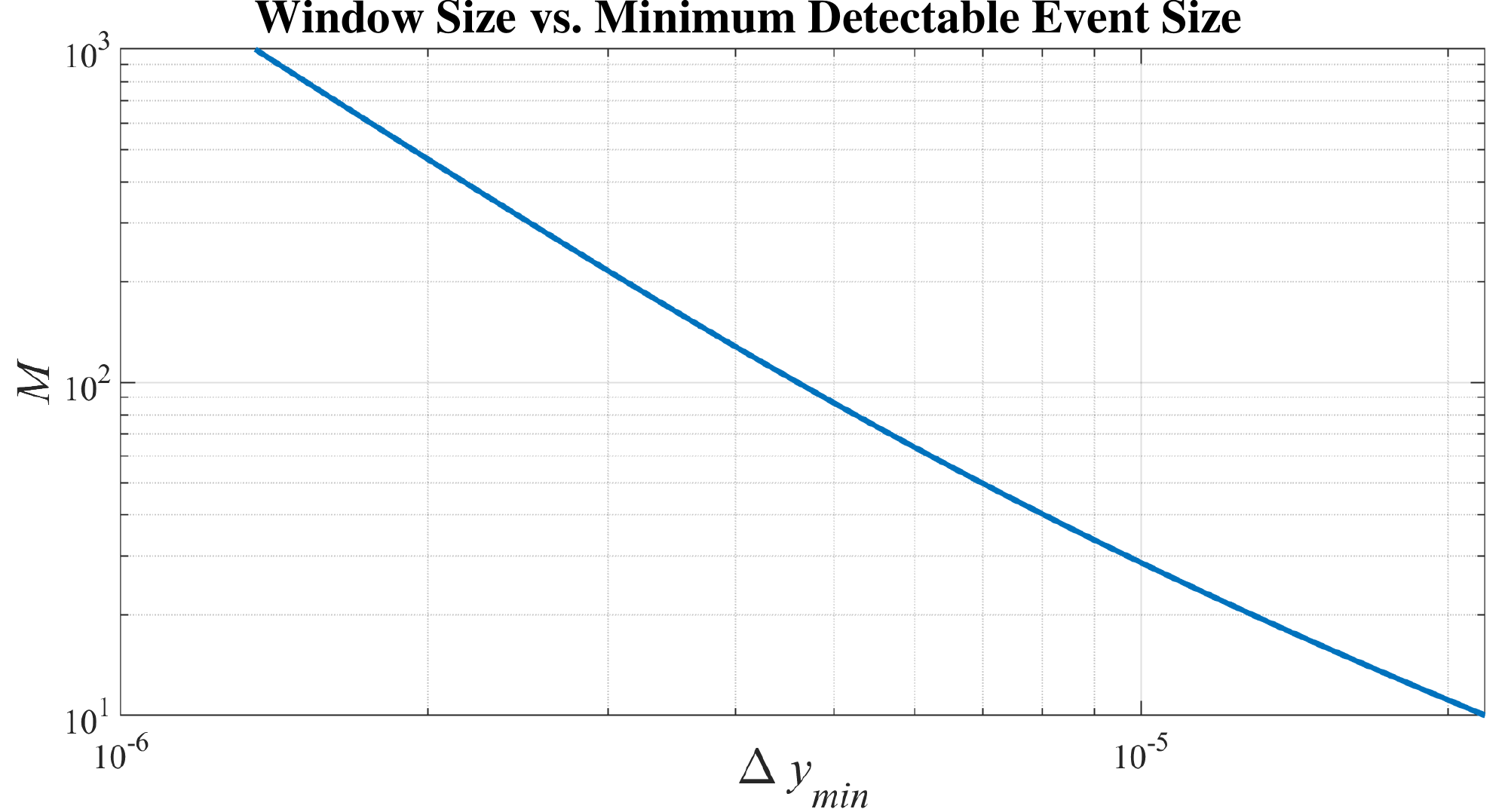}
    \caption{  Window size $M$ versus minimum detectable event size $\Delta{y}_\textsf{\tiny min}$. Plot generated with equations \eqref{eqn:varAKF} and \eqref{eqn:minsize}, and $M$ = $2\, t_e$.}
    \label{fig:ymin_vs_m}
\end{figure}

 The raw sensor output is normalized so that resonance frequency deviation is measured as a fractional (unitless) quantity with respect to the nominal resonance frequency. Time is also normalized in a way that the sampling interval for the sampled sensor output is set to unity. The window duration $M$ is an integer valued quantity that represents the number of samples in the window. For all of the results we present in this paper, we use typical resonator and noise parameters where thermomechanical noise is well resolved above the detection noise floor~\cite{demir2020sensortradeoffs,sadeghi2020frequency}. The inherent response time constant for the raw sensor output with a high quality resonator is set to $\approx 10^5$ samples. The effective bandwidth of the transduction/detection device is chosen to be ten times larger than the resonator linewidth~\cite{demirAKF2021}.

 We consider two values for the minimum event size $\Delta{y}_\textsf{\tiny min}$, {\it i.e.}, $7.5\times10^{-6}$ and $1.25\times10^{-5}$. We use \eqref{eqn:varAKF} and \eqref{eqn:minsize} to compute the minimum elapsed time $t_e$ that is needed as described in Section \ref{sec:intro_KF}.
 We set the window size $M$ to be two times  $t_e$ computed as such and obtain $M\approx50$ ($M\approx20$) for a minimum event size of $7.5\times10^{-6}$ ($1.25\times10^{-5}$). A plot of window size $M$ versus minimum event size $\Delta{y}_\textsf{\tiny min}$ is given in Fig.~\ref{fig:ymin_vs_m} for the resonator, noise and detection parameters we use for the results we present. We use the following performance measures in comparing various methods: 

\noindent
    \textbf{Average Run Time} quantifies computational cost for each new streaming sensor output sample. Values are normalized with respect to the run-time of the thresholding based method with event size $\Delta{y}  = 7.5\times10^{-4}$ and window size $M = 50$. 

\noindent
    \textbf{False detection (Miss)} represents the percentage of experiments in which the detected number of events is strictly larger (smaller) than the number of actual events.

\noindent
    \textbf{Detection Latency} $E[k - k_0$] represents the mean (computed over the experiments, $E[\cdot]$ denotes expectation) difference between the time point ($k$) where decision for event occurrence is made, and the time point ($k_0$) (somewhere in the past $M$ samples) at which event is predicted to have occurred. Detection latency can be modulated by varying the confidence parameter discussed previously. 

\noindent
    \textbf{Bias of event location prediction} $E[k_0] - \gamma$ is the difference between the mean  event location prediction and the actual event location ($\gamma$, same in all experiments). In theory, this difference is zero since the location estimator is unbiased. However, the mean is computed over relatively few number of experiments, and may not correspond to the theoretical value.

\noindent
    \textbf{Root  mean square error (RMSE) of event location prediction} $\text{RMSE}[k_0] = \sqrt{E[(k_0-\gamma)^2]}$ quantifies event location prediction accuracy.

\noindent
    \textbf{RMSE of event size prediction at event decision} 
    \[ \text{RMSE}[\Delta \mathbf{\hat{y}}[k]] = \sqrt{E\left[\left(\frac{\Delta \mathbf{\hat{y}}[k]-\Delta\mathbf{y}}{\Delta\mathbf{y}}\right)^2\right]} \] 
    is a measure for the accuracy of event size predictions at time point ($k$) where decision for event occurrence is made. $\Delta\mathbf{y}$ is the actual event size, and $\Delta \mathbf{\hat{y}}$ is the prediction. The error is normalized with respect to the size of the actual event. 

\noindent
    \textbf{RMSE of event size prediction $10\,M$ samples after event decision} $\text{RMSE}[\Delta \mathbf{\hat{y}}[k+10\,M]]$, computed as above, is a measure for the accuracy of event size predictions at time point ($k+10\,M$), ten window lengths after decision for event occurrence. This is relevant only for thresholding and ML aided Kalman filtering, where event size prediction is continually updated. The event size prediction is a one-shot regression look-up in black-box ML. 

\begin{figure}
    \centering
    \includegraphics[width=\columnwidth]{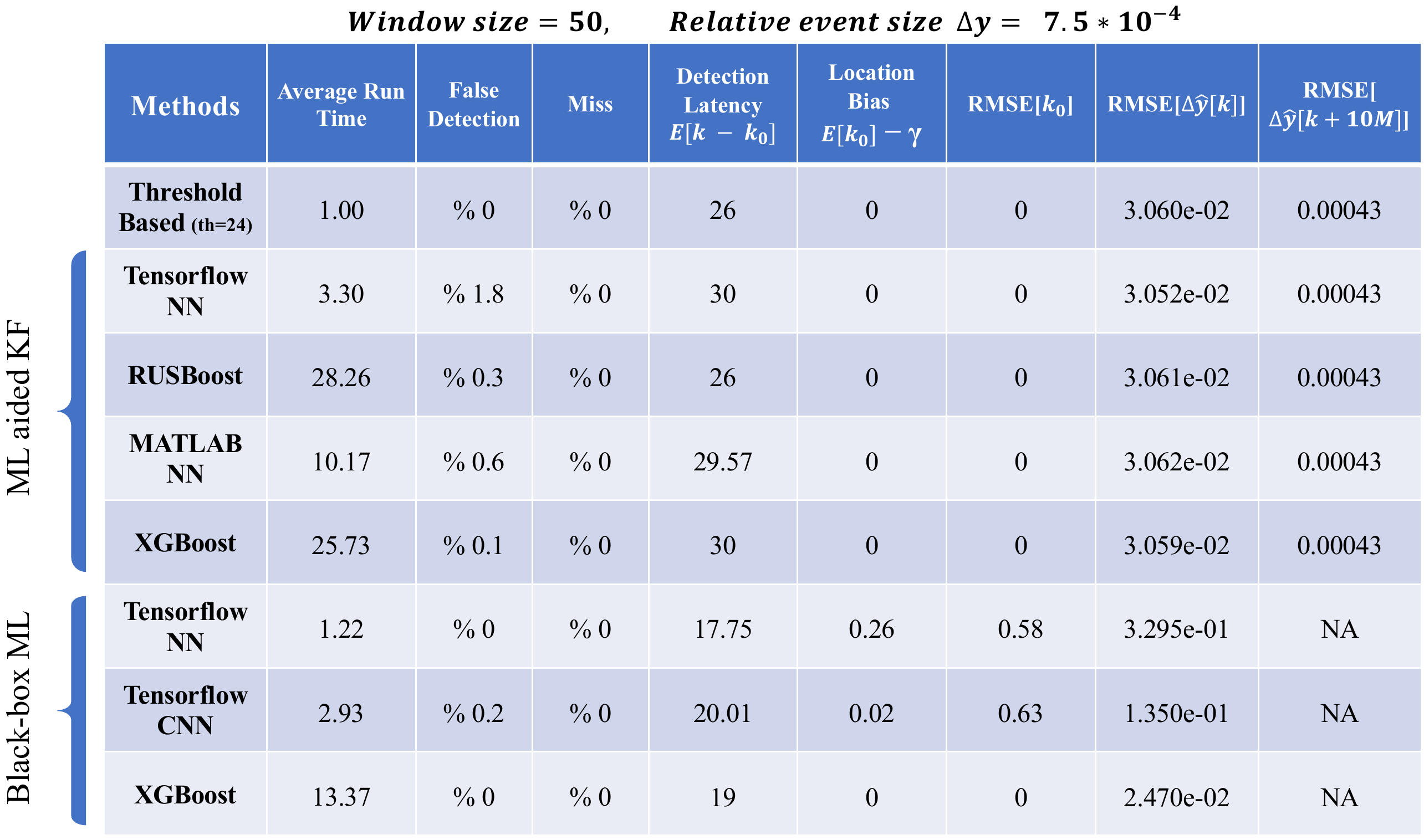}
    \caption{  Event size $\Delta{y}  = 7.5\times10^{-4}$, window size $M = 50$.}
    \label{fig:w50big}
\end{figure}
\begin{figure}
    \centering
    \includegraphics[width=\columnwidth]{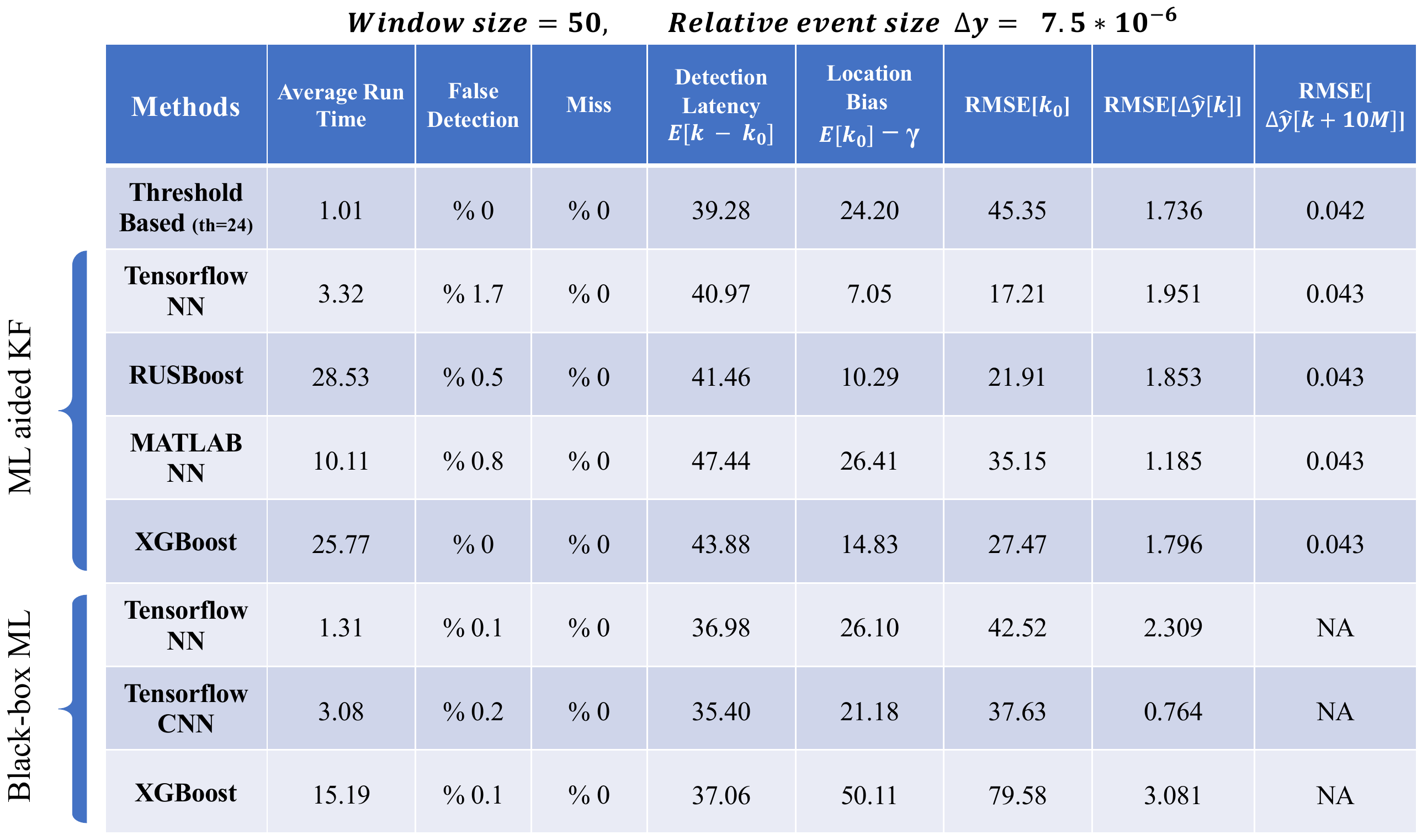}
    \caption{  Event size  $\Delta{y}_\textsf{\tiny min}  = 7.5\times10^{-6}$, window size $M = 50$.}
    \label{fig:w50small}
\end{figure}
\begin{figure}[t]
    \centering
    \includegraphics[width=\columnwidth]{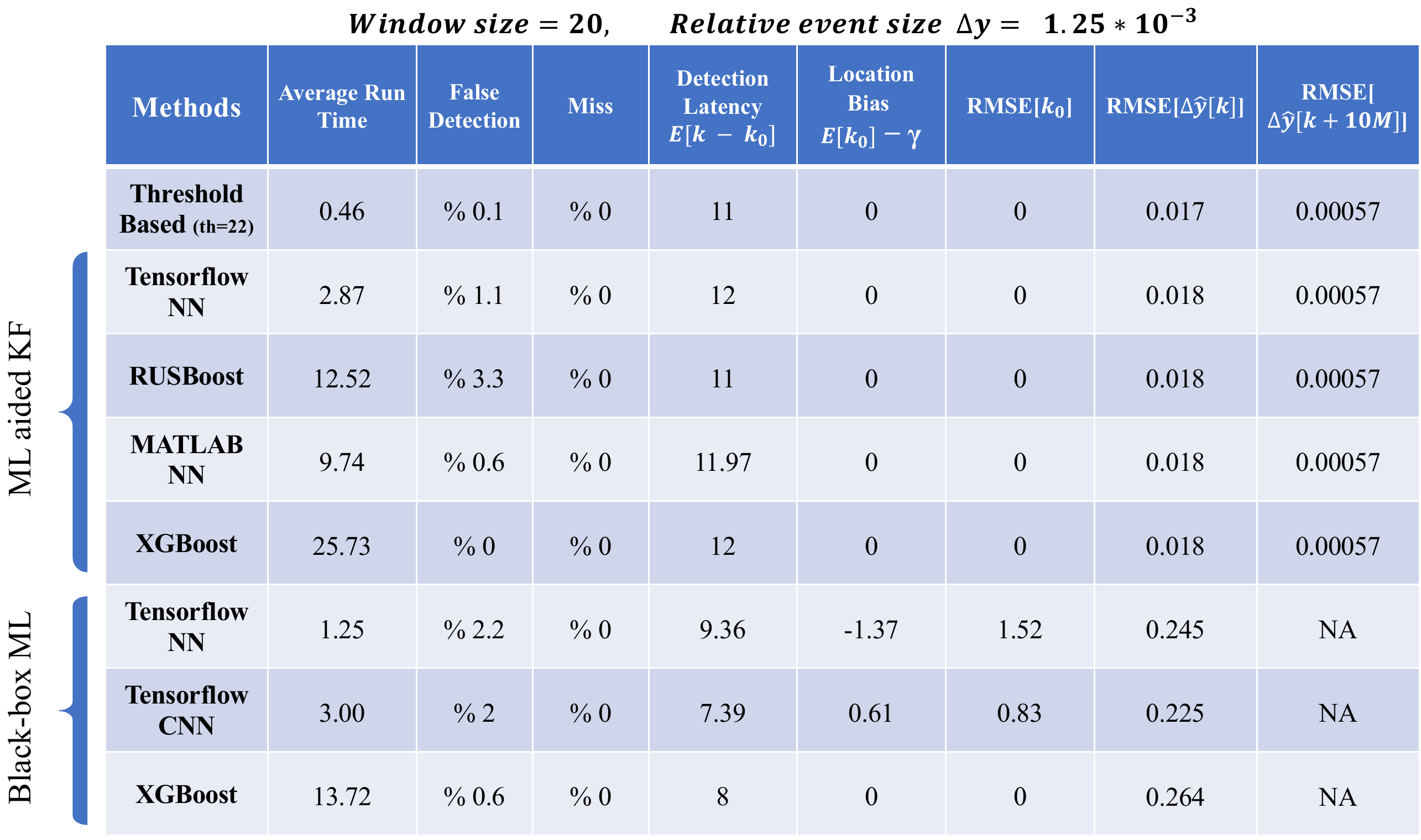}
    \caption{  Event size  $\Delta{y}  = 1.25\times10^{-3}$, window size $M = 20$.}
    \label{fig:w20big}
\end{figure}
\begin{figure}
    \centering
    \includegraphics[width=\columnwidth]{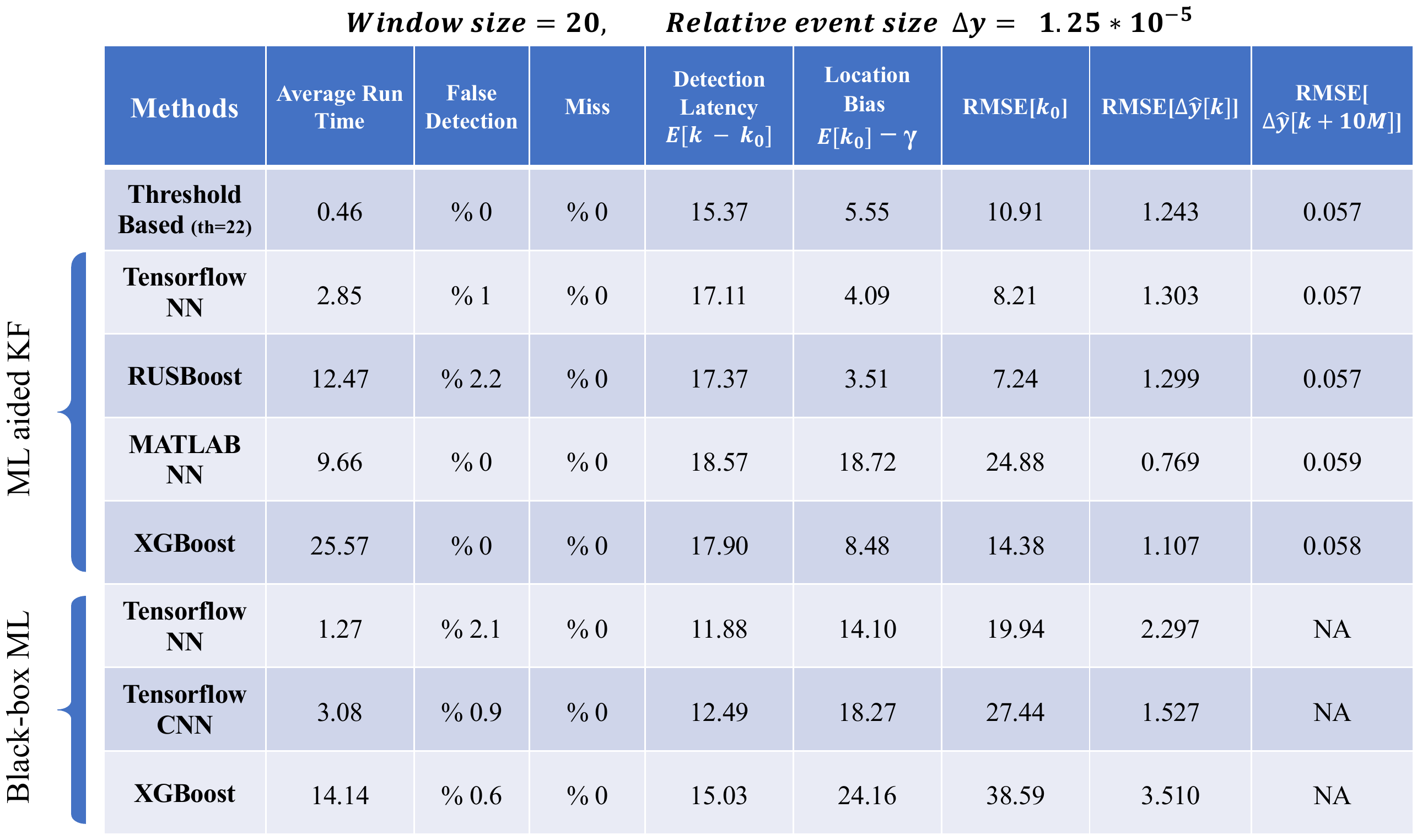}
    \caption{  Event size  $\Delta{y}_\textsf{\tiny min}  = 1.25\times10^{-5}$, window size $M = 20$.}
    \label{fig:w20small}
\end{figure}

We report the results obtained in Figs.~\ref{fig:w50big}, \ref{fig:w50small}, \ref{fig:w20big} and \ref{fig:w20small}.  The rows in the tables correspond to various methods, whereas the performance measures on the columns were defined above. The performance measures were computed over 1000 experiments, where each experiment is a 2000 sample length run with a single event occurring at sample 1000 of the run. We present results for two window sizes, $M=50$ in Figs.~\ref{fig:w50big} and \ref{fig:w50small}, and $M=20$ in Figs.~\ref{fig:w20big} and \ref{fig:w20small}. For each window size, we evaluate performance for the minimum detectable event size that corresponds to the window size considered, as given in Fig.~\ref{fig:ymin_vs_m}. We also present results for events that are two orders of magnitude larger than the minimum. The confidence parameter is chosen as $C=M/2$, relative to the window size $M$. The codes that were used to generate the results in the paper are available at 

{\scriptsize https://github.com/meterdogan07/Nanomechanical-Mass-Spectrometry-ML-KF} 

\subsection{Discussion}

The main disadvantage of the black-box ML approach compared to the likelihood based thresholding and ML aided Kalman filtering schemes is that black box ML technique will fail when inter-event time is smaller than the inherent sensor response time constant, because the training data sessions contain only one event. We could in principle generate training data sessions that also include multiple consecutive events with short time gaps. However, this needs to be done for all event size and inter-event time gap combinations, resulting in a complex and cumbersome data generation and training process. On the other hand, model based Kalman filtering approach can handle multiple consecutive events easily and naturally, since the underlying sensor dynamical model is updated after each event detection so as to capture the cumulative effect of all past events. Thus, training data for the classification networks in the ML aided Kalman filtering technique can contain only one event, as long as no multiple events occur in the same window. The window duration needed is typically much smaller than the inherent response time constant of the resonator. For the results we present in this paper, response time is three orders of magnitude larger. Therefore, using ML on the likelihoods that are processed by the Kalman filter, as opposed to raw sensor data, allows the handling of multiple consecutive events easily. ML is a powerful technique, but it is effective only when the training data and the set of features used in the ML model are judiciously chosen and appropriately processed based on a {\it deep understanding} of the application, as opposed to just using black-box {\it deep learning} on raw data.  

Training is fast for tree-based ML models, but the look-up incurs a high computational cost. This is expected due to the nature of tree-based classification algorithms, where the cost scales exponentially with the depth of the tree. Even though tree-based XGBoost and RUSBoost\cite{RUSboost} techniques both produce accurate results, we conclude that they are not suitable for a DSP/FPGA based implementation intended for a real-time application due to the high look-up cost. The training cost is larger for the neural network based models, but this is acceptable since training is done off-line, only once for a specific sensor in a calibration/setup step before use in a real-time application.

When we compare optimized thresholding with ML aided Kalman filtering, we observe that the robustness and accuracy performance are similar. Event size prediction errors of both are almost equivalent. Optimized thresholding results in almost zero false detections and misses. Threshold based method yields a robust solution where only the threshold parameter needs to be determined, which can be accomplished with our search based method. On the other hand, the behavior of ML based techniques can be controlled only through modifications of the dataset and the training setup. Finally, the cheap computational cost of simple thresholding of likelihoods makes this method preferable, whereas ML aided methods require more complex operations and result in higher run-times.

\section{Conclusions}
\label{sec:concl}
We have thoroughly investigated various ML, maximum-likelihood estimation and Kalman filtering based techniques for mass spectrometry with nanomechanical resonant sensors. We conclude that black-box ML on raw sensor data is inferior when compared with techniques that are based on data processed and refined by a Kalman filter and likelihood computations. Unlike most other recent work on ML techniques in various domains, we conclude that methods that are based on a deep understanding of the application and problem specific techniques that are founded on traditional and well established techniques of detection and estimation theory perform well at a lower computational cost when compared with ML techniques. We plan to implement and test the proposed algorithms on real applications and demonstrate their efficacy experimentally.     
\bibliographystyle{IEEEtran}
\bibliography{references}

\begin{thebibliography}{10}
\providecommand{\url}[1]{#1}
\csname url@samestyle\endcsname
\providecommand{\newblock}{\relax}
\providecommand{\bibinfo}[2]{#2}
\providecommand{\BIBentrySTDinterwordspacing}{\spaceskip=0pt\relax}
\providecommand{\BIBentryALTinterwordstretchfactor}{4}
\providecommand{\BIBentryALTinterwordspacing}{\spaceskip=\fontdimen2\font plus
\BIBentryALTinterwordstretchfactor\fontdimen3\font minus
  \fontdimen4\font\relax}
\providecommand{\BIBforeignlanguage}[2]{{%
\expandafter\ifx\csname l@#1\endcsname\relax
\typeout{** WARNING: IEEEtran.bst: No hyphenation pattern has been}%
\typeout{** loaded for the language `#1'. Using the pattern for}%
\typeout{** the default language instead.}%
\else
\language=\csname l@#1\endcsname
\fi
#2}}
\providecommand{\BIBdecl}{\relax}
\BIBdecl

\bibitem{demir2020sensortradeoffs}
A.~{Demir}, ``Understanding fundamental trade-offs in nanomechanical resonant
  sensors,'' \emph{Journal of Applied Physics}, vol. 129, no.~4, p. 044503,
  2021.

\bibitem{bevsic2023resonance}
H.~Be{\v{s}}i{\'c}, A.~Demir, J.~Steurer, N.~Luhmann, and S.~Schmid,
  ``Resonance frequency tracking schemes for micro-and nanomechanical
  resonators,'' \emph{arXiv:2304.11889 preprint}, 2023.

\bibitem{schmid2023}
S.~Schmid, L.~G. Villanueva, and M.~L. Roukes, \emph{Fundamentals of
  Nanomechanical Resonators}.\hskip 1em plus 0.5em minus 0.4em\relax Springer,
  2023.

\bibitem{demirAKF2021}
A.~Demir, ``Adaptive time-resolved mass spectrometry with nanomechanical
  resonant sensors,'' \emph{IEEE Sensors Journal}, vol.~21, no.~24, pp.
  27\,582--27\,589, December 2021.

\bibitem{Willsky1976}
A.~{Willsky} and H.~{Jones}, ``A generalized likelihood ratio approach to the
  detection and estimation of jumps in linear systems,'' \emph{IEEE
  Transactions on Automatic Control}, vol.~21, no.~1, pp. 108--112, 1976.

\bibitem{chow1976analytical}
E.~Y. Chow, ``Analytical studies of the generalized likelihood ratio technique
  for failure detection,'' \emph{M.S.~thesis, Massachusetts Institute of
  Technology}, 1976.

\bibitem{karniadakis2021physics}
G.~E. Karniadakis, I.~G. Kevrekidis, L.~Lu, P.~Perdikaris, S.~Wang, and
  L.~Yang, ``Physics-informed machine learning,'' \emph{Nature Reviews
  Physics}, vol.~3, no.~6, pp. 422--440, 2021.

\bibitem{multi-output-regression}
L.~Schmid, A.~Gerharz, A.~Groll, and M.~Pauly, ``Tree-based ensembles for
  multi-output regression: Comparing multivariate approaches with separate
  univariate ones,'' \emph{Comput. Stat. Data Anal.}, vol. 179, p. 107628,
  2023.

\bibitem{multi-output-regression2}
H.~Borchani, G.~Varando, C.~Bielza, and P.~Larra{\~{n}}aga, ``A survey on
  multi-output regression,'' \emph{WIREs Data Mining Knowl. Discov.}, vol.~5,
  no.~5, pp. 216--233, 2015.

\bibitem{XGBoost}
T.~Chen and C.~Guestrin, ``{XGBoost},'' in \emph{Proceedings of the 22nd {ACM}
  {SIGKDD} International Conference on Knowledge Discovery and Data
  Mining}.\hskip 1em plus 0.5em minus 0.4em\relax {ACM}, Aug 2016.

\bibitem{CNN}
S.~Kiranyaz, O.~Avci, O.~Abdeljaber, T.~Ince, M.~Gabbouj, and D.~J. Inman,
  ``{1D} convolutional neural networks and applications: A survey,''
  \emph{Mechanical Systems and Signal Processing}, vol. 151, p. 107398, 2021.

\bibitem{RUSboost}
C.~Seiffert, T.~M. Khoshgoftaar, J.~Van~Hulse, and A.~Napolitano, ``Rusboost:
  Improving classification performance when training data is skewed,'' in
  \emph{2008 19th International Conference on Pattern Recognition}, 2008, pp.
  1--4.

\bibitem{sadeghi2020frequency}
P.~Sadeghi, A.~Demir, L.~G. Villanueva, H.~K\"ahler, and S.~Schmid, ``Frequency
  fluctuations in nanomechanical silicon nitride string resonators,''
  \emph{Physical Review B}, vol. 102, p. 214106, Dec 2020.

\bibitem{adamoptimizer}
D.~P. Kingma and J.~Ba, ``Adam: {A} method for stochastic optimization,'' in
  \emph{3rd International Conference on Learning Representations, {ICLR} 2015,
  San Diego, CA, USA, May 7-9, 2015, Conference Track Proceedings}, Y.~Bengio
  and Y.~LeCun, Eds., 2015.

\bibitem{scikit}
F.~Pedregosa, G.~Varoquaux, A.~Gramfort, V.~Michel, B.~Thirion, O.~Grisel,
  M.~Blondel, P.~Prettenhofer, R.~Weiss, V.~Dubourg, J.~Vanderplas, A.~Passos,
  D.~Cournapeau, M.~Brucher, M.~Perrot, and E.~Duchesnay, ``Scikit-learn:
  Machine learning in {P}ython,'' \emph{Journal of Machine Learning Research},
  vol.~12, pp. 2825--2830, 2011.

\bibitem{matlab}
\BIBentryALTinterwordspacing
{The MathWorks, Inc.}, ``{MATLAB\textsuperscript{\textregistered}},'' 2022,
  version 2022b. [Online]. Available:
  \url{http://www.mathworks.com/products/matlab}
\BIBentrySTDinterwordspacing

\end{thebibliography}

\begin{IEEEbiographynophoto}
{Mete Erdogan} is an undergraduate majoring in electrical engineering and computer science at Ko\c{c} University, Istanbul, Turkey.
\end{IEEEbiographynophoto}
\begin{IEEEbiographynophoto}
{Nuri Berke Baytekin} is an undergraduate majoring in electrical engineering and mathematics at Ko\c{c} University, Istanbul, Turkey.
\end{IEEEbiographynophoto}
\begin{IEEEbiographynophoto}
{Serhat Emre Coban} is an undergraduate majoring in electrical engineering and mathematics at Ko\c{c} University, Istanbul, Turkey.
\end{IEEEbiographynophoto}
\begin{IEEEbiographynophoto}
{Alper Demir}
received a B.S. from Bilkent University, Ankara,
Turkey, in 1991, and a Ph.D. from the University of
California, Berkeley, CA, USA, in 1997.  He was a member of technical staff at
Bell Laboratories, Murray Hill, NJ, USA. He has been with 
Ko\c{c} University, Istanbul, Turkey, since 2002.  He received 
the 2002 Best of ICCAD Award, the 2003/2014 IEEE/ACM
William J. McCalla ICCAD best paper awards, and the 2004 IEEE
Guillemin-Cauer Award. 
\end{IEEEbiographynophoto}

\end{document}